\documentclass[conference]{IEEEtran}
\IEEEoverridecommandlockouts
\usepackage{algorithm}
\usepackage{array}
\usepackage{textcomp}
\usepackage{stfloats}
\usepackage{url}
\usepackage{verbatim}
\usepackage{graphicx}
\usepackage{cite}
\usepackage{hyperref}
\usepackage{pifont} 
\usepackage{subcaption}
\usepackage{cite}
\usepackage{amsmath,amssymb,amsfonts}
\usepackage{algorithm}
\usepackage{algpseudocode}
\usepackage{amsmath}
\makeatletter
\renewcommand{\ALG@name}{Algorithm}
\newcommand{\Input}{\item[\textbf{Input:}]}
\newcommand{\Output}{\item[\textbf{Output:}]}
\makeatother
\begin{document}

\title{Large Language Model-Based Semantic Communication System for Image Transmission}

\author{\IEEEauthorblockN{Soheyb Ribouh}
\IEEEauthorblockA{\textit{ Univ Rouen Normandie, INSA Rouen Normandie} \\
\textit{Université Le Havre Normandie, Normandie Univ}\\
\textit{LITIS UR 4108, F-76000 Rouen, France} \\
}
\and
\IEEEauthorblockN{Osama Saleem}
\IEEEauthorblockA{\textit{ INSA Rouen Normandie, Univ Rouen Normandie } \\
\textit{Université Le Havre Normandie, Normandie Univ}\\
\textit{LITIS UR 4108, F-76000 Rouen, France} \\
}
\thanks{}
\thanks{}}



\maketitle

\begin{abstract}
The remarkable success of Large Language Models (LLMs) in understanding and generating various data types, such as images and text, has demonstrated their ability to process and extract semantic information across diverse domains. This transformative capability lays the foundation for semantic communications, enabling highly efficient and intelligent communication systems. In this work, we present a novel OFDM-based semantic communication framework for image transmission. We propose an innovative semantic encoder design that leverages the ability of LLMs to extract the meaning of transmitted data rather than focusing on its raw representation. On the receiver side, we design an LLM-based semantic decoder capable of comprehending context and generating the most appropriate representation to fit the given context.
We evaluate our proposed system under different scenarios, including Urban Macro-cell environments with varying speed ranges. The evaluation metrics demonstrate that our proposed system reduces the data size  4250 times, while achieving a higher data rate compared to conventional communication methods. This approach offers a robust and scalable solution to unlock the full potential of 6G connectivity.

\end{abstract}


\section{Introduction}
Future 6G communications and beyond are expected to revolutionize wireless connectivity by providing ultra-low latency and high reliability, which will significantly advance technologies such as autonomous driving, virtual reality, and remote surgery \cite{saad2019vision}. Motivated by the substantial progress made by Artificial Intelligence (AI) in fields like computer vision, virtual reality, and robotics, the wireless communications community has begun exploring deep learning models to address wireless challenges \cite{wang2023road} \cite{ribouh2022vehicular}. This exploration is strongly supported by the 3GPP consortium, which has confirmed the integration of AI in upcoming releases \cite{3GPP}.

The initial investigations have shown promising results in addressing physical layer tasks, such as channel estimation, channel decoding, and equalization, outperforming traditional methods \cite{ribouh2020multiple}. However, AI-driven physical layer approaches remain limited because they focus primarily on data-driven models that lack the ability to generalize beyond their training data \cite{chaccour2024less}. In contrast, a new knowledge-driven paradigm known as Semantic Communications (SC) has been introduced, aiming to surpass conventional wireless communications methods. SC enables wireless networks to make proactive, logical decisions based on accumulated knowledge from raw data, achieving superior performance, such as high-rate, low-latency, and high-reliability which are crucial for future 6G and beyond \cite{yang2022semantic}. SC aims to reduce communication overhead by transmitting only relevant semantic information. This process involves extracting knowledge from transmitted data using a semantic encoder, then transmitting it over a wireless channel \cite{ribouh2024semantic}. At the receiver side, the received knowledge is utilized by a semantic decoder to either recover the raw data or execute specific tasks in task-oriented communication systems \cite{ribouh2024seecad}. 

SC will revolutionize wireless communication systems by transforming communicated devices into intelligent communicating edge. These new  intelligent nodes will no longer transmit captured data but  understand, interpret, and extract knowledge to be sent. Moreover, these intelligent nodes will have the capability to comprehend and learn from the received knowledge, enabling them to recover data more effectively and make decisions based on that knowledge. This will lead to the built of efficient and intelligent wireless networks.

Semantic communication will also bridge the gap of transmitting large data by optimizing content, which will provide a balance between  data rate and the use of the bandwidth and the spectrum resources. 

Recently, there has been growing attention from researchers in both academia and industry toward semantic communications, as it is expected to revolutionize wireless connectivity by introducing greater intelligence to networks. As a result, the wireless research community has increasingly focused on exploring this area.
\\In \cite{jiang2024large} the authors propose a semantic encoder framework for image transmission. theirs system is designed based on large AI model, where they have used Segment Any Things (SAM) architecture to extract semantic meaning from image data followed by an Adaptive Semantic Compression (ASC) encoding technique to eliminate redundant information within extracted semantic features. 
\\The authors in \cite{nam2024language} introduce a  framework for language-oriented semantic communication. This approach enables machines to communicate using human language messages, which are interpreted and processed through natural language processing (NLP) techniques. The proposed framework include a semantic source coding, which compresses a text prompt into its key headwords at the transmitter and  a semantic knowledge distillation technique for semantic decoding at the receiver, which generates customized prompts by learning the language style through in-context learning.
\\A  Scene Graph-based Generative Semantic Communication (SG2SC) framework is introduced in \cite{yang2024sg2sc}, where they propose a semantic encoder that extract semantic meaning from images in the form of scene graph structure, where a conditional diffusion model is applied for semantic decoding. 
\\A semantic communication model based on Graph Neural Networks (GNN) for task oriented communications is proposed in \cite{zheng2024genet}. In this approach they firstly transform the image to a graph structure then  they use a GNN-based encoder to extract semantic information from this graphs at the transmitter. At the receiver side, an other GNN-based decoder is used to reconstruct the recover semantic graphs to be used for a desired task.
\\In \cite{tong2021federated}, a real-time audio semantic communication system was developed to handle large volumes of audio data. This system employs a Federated Learning mechanism to improve audio signal recovery, achieving efficient convergence and reducing the mean squared error of transmitted audio data.
\\The authors proposed L-DeepSC approach, a lite distributed semantic communication system designed for text transmission \cite{xie2020lite}. This method incorporates deep learning techniques along with channel state information (CSI) aided training to mitigate the impact of fading channels. 

\begin{figure*}[t]
    \centering
    \includegraphics[width=1\textwidth,height=0.24\textheight]{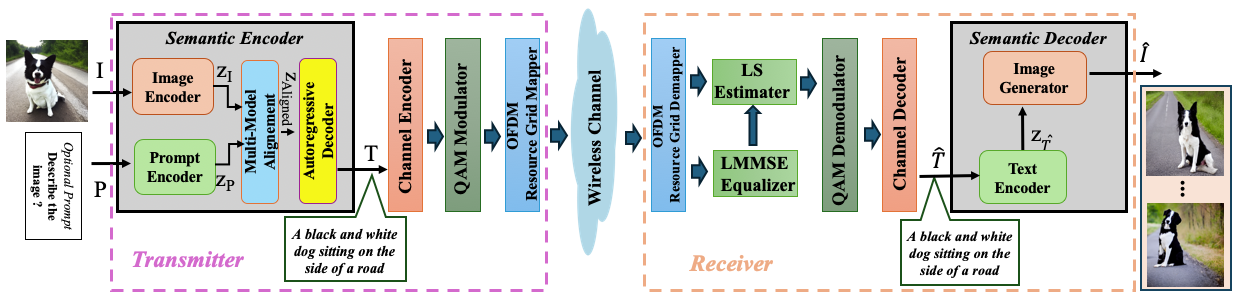}
    \caption{System model}
    \label{fig1}
\end{figure*}

For speech signals, a semantic communication system was presented in \cite{weng2021semantic}. This system utilizes an attention mechanism architecture based on the squeeze-and-excitation (SE) network. The model is designed to be adaptable across various AWGN channel conditions, making it suitable for practical multimedia transmission systems. 
\\A multiple access semantic communication model based on deep learning, named DeepMA, was proposed in \cite{zhang2023deepma}. This approach integrates joint source-channel coding using a neural network with an encoder-decoder architecture. It was deigned for a multiple access wireless image transmission task.
\\ \cite{lokumarambage2023wireless}, the authors introduced a semantic communication model for image transmission considering the challenges of a noisy channel. This system employs semantic segmentation extraction as an encoder at the transmitter, and a GAN network as a semantic decoder at the receiver to reconstruct the transmitted image. 
The aforementioned research works have been built upon a baseline wireless communication system that includes only a semantic encoder/decoder at the transmitter and a channel encoder/decoder at the receiver. Furthermore, these systems have been tested using the AWGN channel model, which is less challenging and not fully representative of real-world wireless channels. In contrast, this work proposes an OFDM-based semantic communication framework that incorporates all the necessary components of a wireless communication system. We introduce a new semantic encoder block designed using a Large Language Model (LLM) to extract information from the source data and a semantic decoder based on a generative diffusion model to accurately recover the source data. 
\\The main contributions of this paper are summarized below:
\begin{enumerate}
    \item We introduce a comprehensive end-to-end semantic wireless communication system based  LLMs. We propose an novel semantic encoder design that leverages LLMs to extract the meaning from raw data. On the receiver side, we design an LLM-based semantic decoder capable of comprehending context and generating the most appropriate representation to fit the given context.
    \item We evaluate our system over an Urban Macro-cell (UMa) wireless channel, as defined by 3GPP consortium in \cite{c19}, demonstrating its efficiency compared to conventional system
\end{enumerate}

\section{System Model}
\subsection{Semantic communication background}
Some research works have explored semantic theory, where most of them are based on logical probability, which follows the framework of conventional information theory. However, it remains uncertain whether this approach can effectively quantify semantic communications and it still an active research area with significant potential for further exploration in the future \cite{qin2021semantic}. 
\\Based on this approach semantic communication can be qualified using the following concepts : \begin{itemize}
    \item \textbf{Semantic entropy}
   Several definitions have been introduced to measure the semantic information. Building on these, and aligned with by Carnap and Bar-Hillel work \cite{carnap1952outline} that define the logical probability of a  message $x$ as:
\begin{equation}
   L_{p}(x)= \frac{P(W_x)}
   {P(W)} = \frac{\underset{w \in W, w \models x}\sum p(w)}{\underset{w \in W}\sum p(w)}
\end{equation}
Where $W$ represent the semantic representation or meaning space of the message $x$ and $w \models x$ refer to the space where  $w$ satisfies or is consistent with the message $x$. $P(W_s)$ refer to the likelihood of the message s being true across the subset of model that align with its meaning. It quantifies the informativeness of $s$ by measuring how much of the meaning space it occupies. $P(W)$ is the total likelihood of all meaning space of the message $x$.In a normalized  the probability $P(W)$ is equal to 1 ($\underset{w \in W}\sum P(w)=1$). Thus the logical probability can be simplified to :
\begin{equation}
     L_{p}(x) = \underset{w \in W, w \models x}\sum p(w)
\end{equation}
\\The semantic entropy $H_se$ can be described as follows:
\begin{equation}
   H_{se}(x)=-\log(L_{p}(x))
\end{equation}

    \item \textbf{Semantic channel capacity}
In the context of semantic transmission the semantic channel capacity is expressed as \cite{qin2021semantic} :  
\begin{equation} \label{eq_c}
        C_{se} = \underset{p(Z|X)}{\sup} \{I(X;\hat{X}) - H(Z|X) + H_{se}(\hat{X})\}
\end{equation}
Where $I(X;\hat{X})$ is the mutual information between the raw data to transmitter $X$ and the reconstructed data $\hat{X}$ at the receiver side. $p(Z|X)$ represents the conditional probabilistic distribution that refer to  the semantic coding function.
 \\ $H(Z|X)$ represents the conditional entropy of the features $Z$ given the data $X$. It gives the uncertainty of $Z$ when you have a given value of $X$. $H(Z|X)$ characterizes the semantic encoding noise. 
\\ $H_{se}(\hat{X})$ is the entropy of $\hat{X}$ that measure the uncertainty associated to the recontacted data $\hat{X}$. A higher  $H_{se}(\hat{X})$ enhances the receiver's ability to recontract the data correctly.
\\From Equation \ref{eq_c}, we deduce that in semantic transmission we can handle two hypotheses:
\begin{enumerate}
    \item If the semantic noise $H(Z|X)$ is higher than $H_{se}(\hat{X})$, the receiver cannot overcame this semantic ambiguity and the semantic channel capacity $C_{se}$ is lower than Shannon capacity, which results in an incorrect generated data . 
    \item  If $H_{se}$  is higher than $H(Z|X)$, this means that the receiver is able to deal with the semantic noise and can generate the data correctly.
\end{enumerate}  
\end{itemize} 
\subsection{The proposed system}
In this section, we propose a wireless communication pipeline designed for transmitting high-level semantic information of image data over a wireless channel. As shown in Figure \ref{fig1},  our system comprises two main components that serve as the core elements enabling semantic communication: the semantic encoder at the transmitter and the semantic decoder at the receiver. Each component is described in detail below.
\begin{itemize}
    \item \textbf{Semantic Encoder:}  It is responsible for extracting the meaning and knowledge from the raw images resulting a compact representation to minimize transmission overhead while retaining the relevant information. As shown in Algorithm \ref{Algo_SE}, the semantic encoder will ensure the mapping of the input image $I \in \mathbb{R}^{H \times W \times C} $ to a knowledge text sequence output $T = \{t_1, t_2, \dots, t_n\}$. \\The Semantic encoding process begins by encoding the input image I into latent features using an image encoder function. If a prompt is provided, it is encoded into text embeddings in the second step. In the next step, the  image and text features are aligned into a unified multimodal space. Then, it generate the text token by token using an autoregressive decoding function. Finally, the output the generated knowledge text T is provided.
\begin{algorithm}[t]
\caption{Semantic Encoder}\label{Algo_SE}
\begin{algorithmic}[1]
\Input Image : $I \in \mathbb{R}^{H \times W \times C}$,  \par \textit{Optional  prompt : $P = \{p_1, p_2, \dots, p_m\}$}
\Output Generated text $T = \{t_1, t_2, \dots, t_n\}$
\State Compute image features : $\mathbf{z}_I \gets En_I(I)$ 
\If {($P$ = True)}
    \State Compute prompt embeddings : $\mathbf{z}_P \gets \text{$En_p$}(P)$ 
\Else
    \State Set $\mathbf{z}_P \gets \emptyset$
\EndIf
\State Fuse image and text features: $\mathbf{z}_{\text{aligned}} \gets \text{Align}(\mathbf{z}_I, \mathbf{z}_P)$
\State Initialize the text sequence: $t_0 \gets [ \text{START}] $.
\For{$i = 1$ to $n$}
    \State Generate next token: $t_i \gets \text{$Auto_{reg}$}(t_{<i}, \mathbf{z}_{\text{aligned}})$
    \If{($t_i = [\text{END}])$}
        \State Break
    \EndIf
\EndFor
\State $T \gets \{t_1, t_2, \dots, t_n\}$
\end{algorithmic}
\end{algorithm}

    \item \textbf{Semantic Decoder:} It is responsible for interpreting the recover meaning aiming to reproduce an image that closely resembles the original transmitted raw image while maintaining the semantic integrity. As shown in \ref{Algo_SD}, the semantic decoding  process begins by computing a latent representation \( \mathbf{z}_{\hat{T}} \) of the input estimated text knowledge sequence \( \hat{T} \) using a text encoding function \( f_z \). This latent representation, which captures the semantic meaning of the text, is then passed to a generative function \( f_{\text{gen}} \) to produce the corresponding image \( \mathbf{\hat{I}} \).  As the final result, it will provide the generated image that effectively bridging text-based semantic knowledge  \( \hat{T} \) to visual  image representation \( \mathbf{\hat{I}} \).

\begin{algorithm}[t]
\caption{Semantic Decoder} \label{Algo_SD}
\begin{algorithmic}[1]
\Input Estimated Text knowledge sequence \par $\hat{T} \gets \{\hat{t_1}, \hat{t_2}, \dots, \hat{t_n}$\}.
\Output Generated image \( \mathbf{\hat{I}} \).

\State Compute the text latent representation:
\[
\mathbf{z}_{\hat{T}} \gets f_{z}(\hat{T})
\]

\State Generate the image :
\[
\mathbf{\hat{I}} \gets f_{\text{gen}}(\mathbf{z}_{\hat{T}})
\]

\State \textbf{Output}: Generated image \( \mathbf{\hat{I}} \).

\end{algorithmic}
\end{algorithm}

\end{itemize}
\section{Experimental setup}
To evaluate the performance of our proposed system, we implemented it using the Sionna library \cite{c20} in a simulated 6G cellular communication setting, where a user equipment (UE) transmits information to a base station (BS) within an urban environment. A detailed overview of the experimental settings is provided in Table \ref{table1}. The implementation has been carried out across two levels as follows : \\
\textbf{At the semantic level : } the semantic encoder is built upon Large Language and Vision Assistant (LLaVA) model \cite{liu2024visual}, integrating advanced components for efficient multimodal processing. For image encoding, it utilizes a Vision Transformer (ViT)-based architecture\cite{dosovitskiy2020image}, which extracts latent features from the input image. For text encoding, the system employs a fine-tuned LLaMA (Large Language Model Meta AI) \cite{touvron2023llama}, a transformer-based architecture that processes tokenized text into rich, contextual embeddings. The text and image features are aligned in a unified multimodal space using a shared projection mechanism, enabling coherent and semantically relevant output. This LLaVA model ensures the accurate transformation of multimodal inputs into token-by-token text generation through an autoregressive decoder, provinding a  text based on the image's knowledge. \\At the receiver side, the semantic decoder implementation, we used a combination of the text encoder from CLIP model\cite{radford2021learning} for semantic understanding and Stable Diffusion Model \cite{rombach2022high} for high-quality image generation. The CLIP text encoder architecture is based on the Transformer, which converts tokenized text into a fixed-dimensional semantic latent vector. This latent space representation serves as a compact and semantically rich descriptor of the input text. The stable diffusion pipeline  leverage Denoising Diffusion Probabilistic Models (DDPM)\cite{ho2020denoising} to iteratively generate high-quality images. It incorporates cross-attention mechanisms to directly condition the generation process on the text embeddings, ensuring that the output images align semantically with the input descriptions.\\
\textbf{At the channel level :} as shown in figure \ref{fig1} the UE's signal is first encoded using a Low-Density Parity Check (LDPC) encoder. The encoded signal is then modulated with Quadrature Amplitude Modulation (QAM) to generate baseband symbols. These symbols are mapped onto an Orthogonal Frequency Division Multiplexing (OFDM) resource grid, with known pilot symbols embedded for channel estimation. This resource grid is equipped with 128 subcarriers and a subcarrier spacing of 240 kHz . It is structured with 14 OFDM symbols per frame, utilizing a Kronecker pilot pattern to enhance channel estimation accuracy. Each grid carries 4-QAM-modulated data, encoding 2 bits per symbol. Once transmitted by by the UE , the resource grid passes through an Urban Macrocell (UMa) channel as specified by 3GPP \cite{c19}. At the receiver, the resource grid undergoes demapping, followed by channel estimation using the known pilot symbols. The signal is then equalized to mitigate multipath fading and inter-symbol interference (ISI). Next, the equalized signal is demodulated to generate log-likelihood ratios (LLRs), which are passed through an LDPC decoder to corrects errors caused by the wireless channel and removes the redundancy introduced by the channel encoder to recover the semantic information.

\begin{table}[t!]
    \centering
    \caption{Wireless Communication Parameters}
    \begin{tabular}{|c|c|} \hline
        \textbf{Parameter} & \textbf{Value} \\ \hline 
        Carrier Frequency & 28GHz \\ \hline

        Code rate & 0.5 \\ \hline
        No. of Subcarriers & 128 \\ \hline
        Subcarrier Spacing & 240KHz \\ \hline
        No. of Transmitter Antenna & 1 \\ \hline
        No. of Receiver Antenna & 2 \\ \hline
        No. of OFDM symbol & 14 \\ \hline
        Physical Channel & UMa \\ \hline
        Speed range & 60-120 km/h \\ \hline
    \end{tabular}
    \label{table1}
\end{table}

\section{Results and Discussion}

\begin{figure}[t]
  \centering
  \includegraphics[width=0.45\textwidth,height=0.25\textheight]{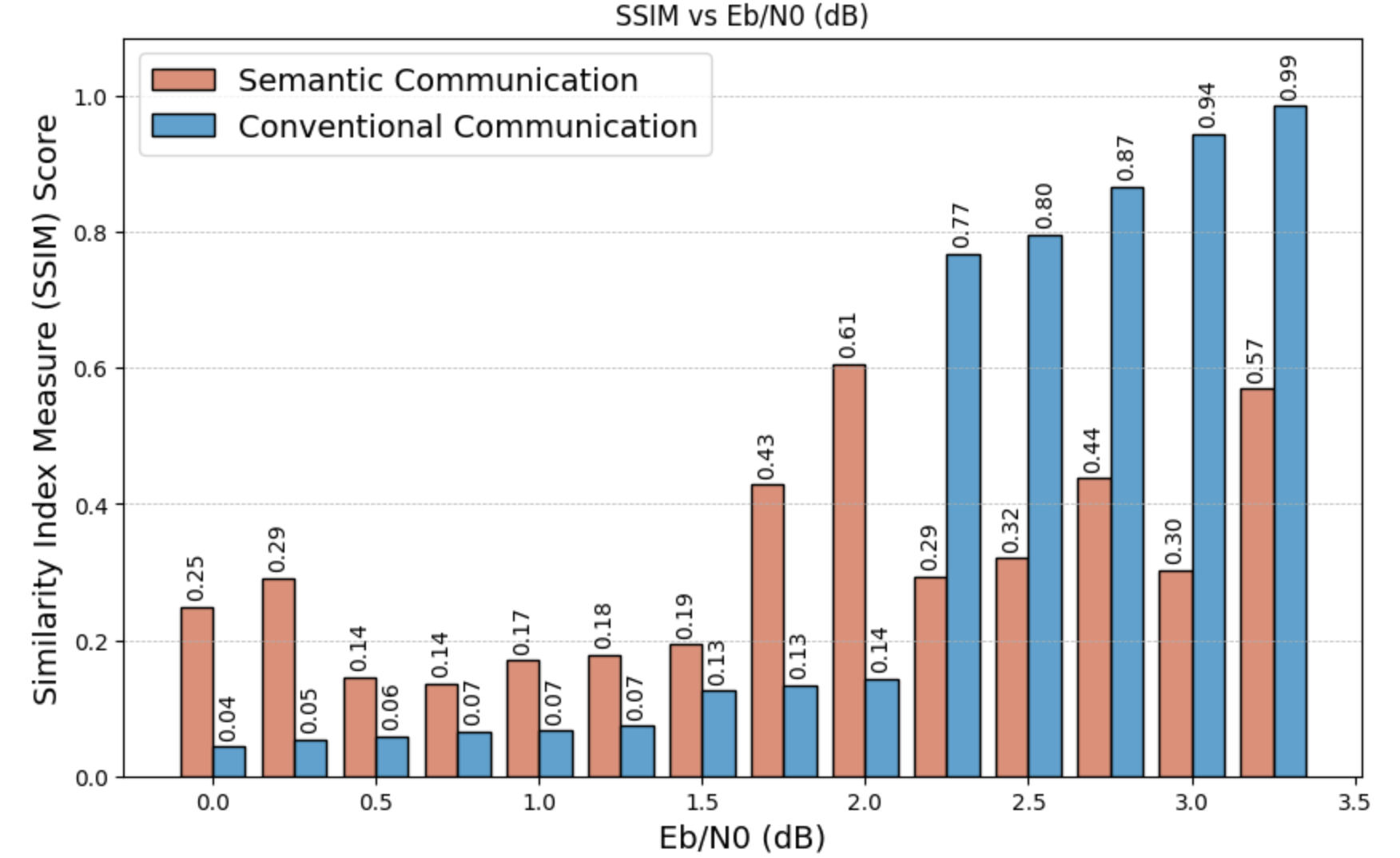}
  \caption{SSIM vs. SNR for semantic and conventional communication.}
  \label{fig5}
\end{figure}

\begin{table}[htb]
    \centering
    \caption{Effective Data Rate}
    \begin{tabular}{|c|c|c|} \hline
        \textbf{Algorithm} & \textbf{Data Size} & \textbf{Data Rate} \\ \hline
        Semantic Communication & 1.392 Kb & 23.05 Mbps \\ \hline
        Conventional Communication & 5.916 Mb & 19.34 Mbps \\ \hline
    \end{tabular}  
    \label{table2}
\end{table}

To evaluate recontacted image quality of our proposed semantic system  compared  conventional communication, we computed the Structural Similarity Index Measure (SSIM) across various SNR values. The results are shown in Fig.~\ref{fig5} which demonstrate that at lower SNR values, SSIM for semantic communication is superior to that of the conventional communication. This is because, even with altered bits, the semantic decoder is able to  generate the image successfully, whereas missing bits in conventional communication have a greater impact on image quality. However, at higher SNR values, conventional communication yields a higher SSIM. This  is due to the  resemblance of the received image to the transmitted one. In contrast, with semantic communication, there will always be some degree of variation. This variation is not a flaw but an inherent characteristic of semantic communication, as it reconstructs the image based on prior knowledge and understanding of the semantic content, rather than replicating the original data exactly, ensuring efficient and meaningful data transmission.

We calculated the effective data rates for both semantic and conventional communication. 

The effective data rate is computed using the formula: \(\text{Effective Data Rate} = R \times S\), where \(R\) represents the Data Rate and \(S\) represents the Success Rate. The Success Rate is defined as \(S = \frac{B_s}{B_t}\), where \(B_s\) is the Successfully Transmitted Bits and \(B_t\) is the Total Transmitted Bits and Data Rate is calculated as \(R = \frac{B_t}{T}\), where \(T\) is the Total Transmission Time. Furthermore, the Total Transmission Time is calculated as \(T = N \times T_{\text{symbol}}\), where \(N\) is the Number of OFDM Symbols and \(T_{\text{symbol}} = \frac{1}{f_s}\), with \(f_s\) representing the Subcarrier Spacing.

The results are shown in Table~\ref{table2} where it can be seen that semantic communication achieves a higher data rate, with \textbf{23.05 Mbps} compared to \textbf{19.34 Mbps} for conventional communication. This improvement results from semantic communication’s ability to selectively transmit essential information rather than the entire image, enabling efficient data transmission and reducing bandwidth demands. From Table~\ref{table2}, we can observe that semantic communication reduces the transmitted data size to \textbf{$1.392 Kb$}, achieving a compression ratio of \textbf{$4250$}.

\begin{figure}[t]
  \centering
  \includegraphics[width=0.45\textwidth,height=0.25\textheight]{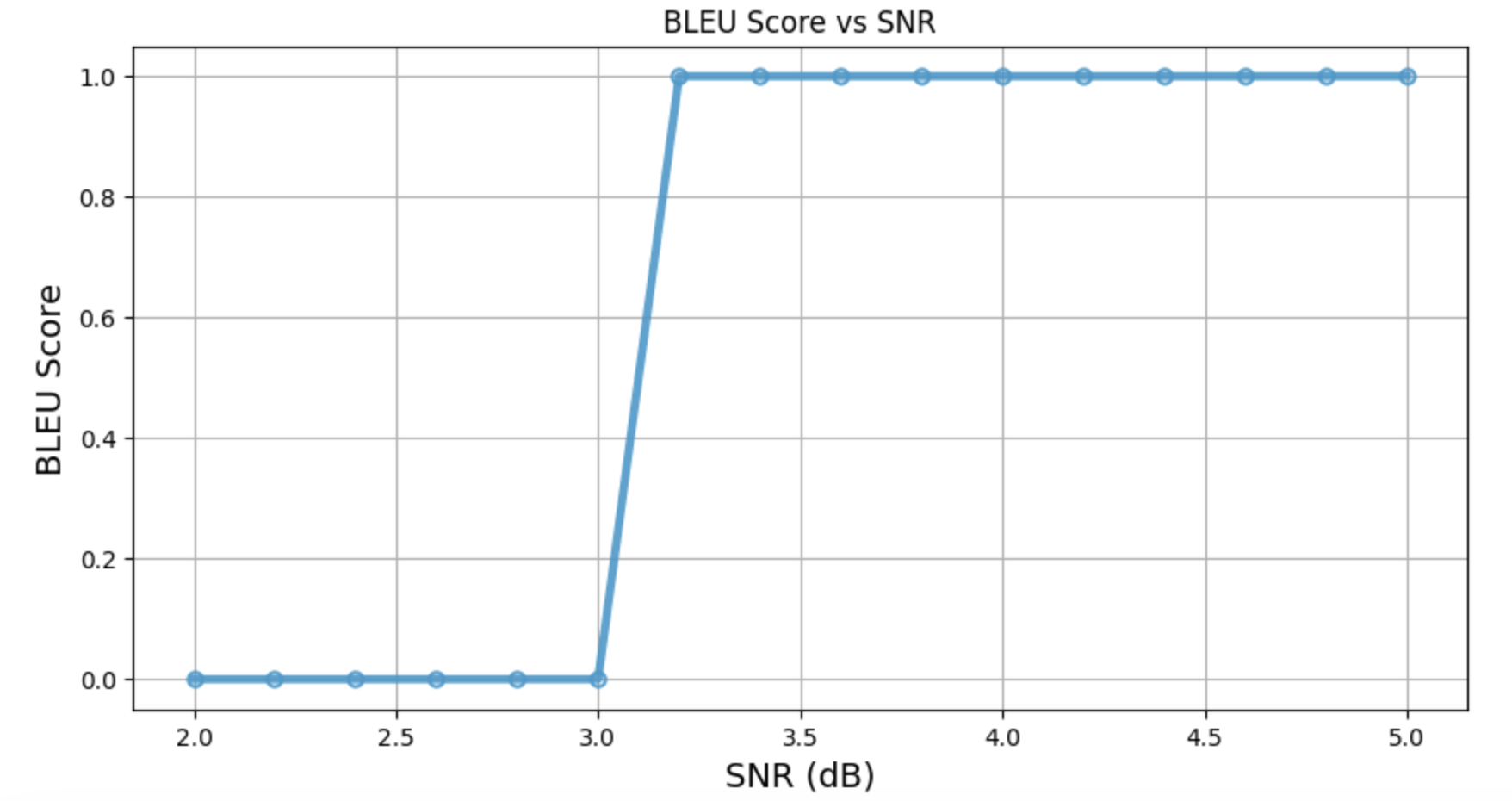}
  \caption{BLEU score vs. SNR for semantic prompts.}
  \label{fig4}
\end{figure}

To assess the performance of our proposed system we compute the Bilingual Evaluation Understudy (BLEU) to evaluate the quality of receiver semantic prompts compared to the original one provided by the semantic encoder. 
Fig.~\ref{fig4} presents the achieved BLUE score over different SNR  value for a semantic prompt derived from an image captured by an UE moving in an urban environment. The figure shows a positive correlation: as the SNR value increases, the BLEU score also improves, reaching its maximum at an SNR of beyond $3.3 dB$.

\begin{figure*}[t]
  \centering
  \includegraphics[width=1.0\textwidth,height=0.25\textheight]{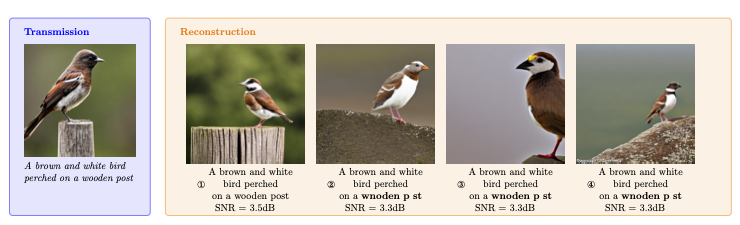}
  \caption{Illustration of the transmission and reconstruction process. The transmitted image and its corresponding semantic representation (prompt) are presented alongside the received prompt and the reconstructed image at the receiver.}
  \label{fig2}
\end{figure*}

Fig.~\ref{fig2} illustrates the transmitted and received image, alongside the corresponding semantic prompt. In this scenario, we assume that the UE is moving with a speed of 90 km/hr, with signal transmitted over UMa channel at Signal-to-Noise Ratio (SNR) of $3.5$dB. The output prompt of the semantic encoder is: \textit{“A brown and white bird perched on a wooden post.”}.On the receiver side, the semantic information prompt was successfully recovered. As a result, the image reconstructed by the semantic decoder closely resembles the transmitted one as shown in Fig.\ref{fig2}\ding{172}.

In the next step, we lowered the SNR to $3.3$dB to test performance, where we observe slightly altered bits. This  results in the following  reconstructed semantic information prompt: \textit{“A brown and white bird perched on a \textbf{wnoden p st}.”} Despite the missing information ("a wooden post."), the semantic decoder still successfully reconstructed an image of a white and brown bird, though without the wooden post, as illustrated in Fig.\ref{fig2}\ding{173}. We processed several iterations through the semantic decoder using this received semantic prompt as input. The results confirm that the semantic decoder successfully reconstructs a brown and white bird in a sitting position as shown in Fig.\ref{fig2}\ding{174} and Fig.\ref{fig2}\ding{175}.

\section{conclusion and future works}
In this paper, we presented a novel semantic communication framework for image transmission. Our approach integrates an LLM-based semantic encoder and decoder into an OFDM framework, enabling efficient context-aware transmission by prioritizing the meaning of the data. The evaluation of the proposed system across various scenarios highlights significant improvements in data reconstruction under low SNR conditions and data compression compared to traditional communication systems. This work validates the potential of LLMs as a transformative solution for designing next-generation wireless communication systems. As a future research direction, we aim to extend this framework to other data modalities and large-scale real-world applications.

\vfill

\end{document}